\documentclass{llncs}
\usepackage[T1]{fontenc} 
\usepackage{graphicx} 
\usepackage[boxed,vlined,ruled,linesnumbered]{algorithm2e}
\usepackage{cite}
\usepackage{amsmath}

\usepackage{color} 
 
\begin{document}

\title{Understanding Branch Cuts of Expressions
\thanks{The final publication is available at http://link.springer.com.}
}
\author{Matthew England \and Russell Bradford \and James H. Davenport \and David Wilson}
\institute{
University of Bath, Bath, BA2 7AY, U.K. \\
\email{ {\tt \{M.England, R.J.Bradford, J.H.Davenport, D.J.Wilson\}@bath.ac.uk}},\\ 
WWW home page: \texttt{http://people.bath.ac.uk/masjhd/Triangular/}
}
\maketitle

\begin{abstract}

We assume some standard choices for the branch cuts of a group of functions and consider the problem of then calculating the branch cuts of expressions involving those functions. Typical examples include the addition formulae for inverse trigonometric functions.  Understanding these cuts is essential for working with the single-valued counterparts, the common approach to encoding multi-valued functions in computer algebra systems.  While the defining choices are usually simple (typically portions of either the real or imaginary axes) the cuts induced by the expression may be surprisingly complicated.  We have made explicit and implemented techniques for calculating the cuts in the computer algebra programme {\sc Maple}. We discuss the issues raised, classifying the different cuts produced. The techniques have been gathered in the \texttt{BranchCuts} package, along with tools for visualising the cuts.  The package is included in {\sc Maple 17} as part of the \texttt{FunctionAdvisor} tool.



\end{abstract}

\section{Introduction} \label{SEC:Intro}

We consider the problem of calculating the branch cuts of expressions in a single complex variable.  When defining multi-valued functions mathematicians have a choice of where to define the branch cuts.  There are standard choices for most well-known functions \cite{AS64, OLBC2010, NIST}, usually following the work of Abramowitz and Stegun.  These choices were justified in \cite{CDJW00} and match the choices within the computer algebra programme {\sc Maple} for all elementary functions except arccot (for reasons explained in \cite{CDJW00}).  Within this paper we assume branch cut definitions matching those of {\sc Maple} (which may be observed using Maple's \verb+FunctionAdvisor+ by giving the function name without an argument).  We note that a different choice would not lead to any fewer or less complicated issues.

Handbooks (including online resources such as \cite{NIST}) and software usually stop at these  static definitions.  However, our thesis is that this knowledge should be dynamic; processed for the user so it is suitable for their situation.  Hence we consider the problems that follow after the initial choice of definition is settled.  This will involve symbolic computation but is also an issue of Mathematical Knowledge Management (following the \emph{process view} of MKM in \cite{CF09}).  

We wish to axiomatically understand the branch cuts of \emph{expressions} in multi-valued functions, such as functions applied to a non-trivial argument, function compositions, and function combinations (sum, product, relations).  
Many of the well-known formulae for elementary functions, such as addition formulae for inverse trigonometric functions, are such expressions.  Care needs to be taken when working with multi-valued functions since there are different, often unstated, viewpoints possible as discussed in \cite{Davenport07, Davenport10}.  Most computer algebra software (and indeed most users) tend to work with multi-valued functions by defining their single-valued counterparts which will have discontinuities over the branch cuts.  As a result, relations true for the multi-valued functions may no longer be true for the single valued counterparts and hence understanding the branch cuts of the relations becomes essential for working with them efficiently.  

Despite the importance of understanding such branch cuts, the authors are not aware of any (available) software which calculates them beyond the original definitions.  It also seems rare for them to get a detailed mathematical study in their own right, beyond their introduction and simple examples, with \cite{Markushevich1965a} one notable exception.

We denote multivalued functions evaluating to sets of values using names with upper cases (i.e. Arctan, Sqrt$(z)$, Log) and denote their single valued counterparts by the normal notation (i.e. $\arctan, \sqrt{z}, \log$).  So, for example, Sqrt$(4)=\{-2,2\}$ while $\sqrt{4}=2$.  (Given our above choice of branch cut definitions, this now means our notation throughout the paper matches the commands in {\sc Maple}.)  We note that when dealing with sets of values for multi-valued functions not all combinations of choices of values of will be meaningful and sometimes the choices for sub-expression values are correlated.  

A simple example of the problem described above is that while the identity
$
\mathrm{Sqrt}(x)\mathrm{Sqrt}(y) = \mathrm{Sqrt}(xy)
$
is true (in the sense that the set of all possible products of entries from the two sets on the right is the same as the set on the left), the single valued counterpart 
$\sqrt{x}\sqrt{y}=\sqrt{xy}$ is not universally true (for example when $x=y=-1$).  The regions of truth and failure are determined by the branch cuts of the functions involved.  

The standard choices for branch cuts of the elementary functions are reasonably simple, always taking portions of either the real or imaginary axes.  Indeed, all the branch cut definitions within {\sc Maple} adhere to this rule (including those from outside the class of elementary functions).  However the branch cuts invoked by the expressions built from these can be far more complicated.  

\begin{figure}[ht] 
\begin{center}
\includegraphics[width=4.5cm]{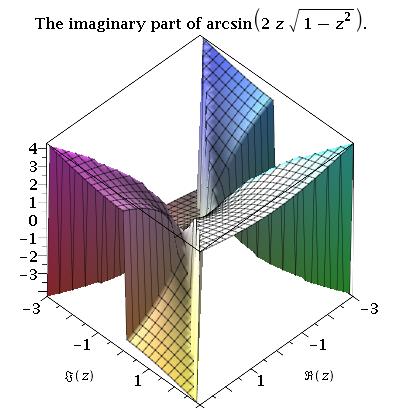}
\hspace*{0.3cm}
\includegraphics[width=4.5cm]{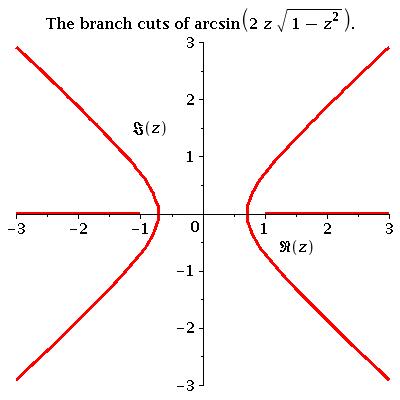}
\end{center}
\caption{Plots relating to $\arcsin(2z\sqrt{1-z^2})$.} 
\label{fig:arcsin}
\end{figure}

Consider for example the composite function $\arcsin(2z\sqrt{1-z^2})$ which is a term from the double angle formula for $\arcsin$.  While $\arcsin(z)$ has simple branch cuts (when $z$ takes values along the real axis, to the left of $-1$ and to the right of $+1$), the branch cuts of the composite function are curves in the complex plane as demonstrated by the plot of the function on the left of Figure \ref{fig:arcsin}.  

The cuts can be described by the four sets below which are visualised in the image on the right of Figure \ref{fig:arcsin}.
\begin{equation}
\label{eq:arcsin}
\begin{array}{rclcrcl}
\big\{\Im(z) &=& 0, 1 < \Re(z)\big\}, & \quad &
\big\{\Im(z) &=& \Im(z), \Re(z) = -(1/2)\sqrt{2+4\Im(z)^2}\big\},  \\
\big\{\Im(z) &=& 0, \Re(z) < -1\big\}, & \quad & 
\big\{\Im(z) &=& \Im(z), \Re(z) = (1/2)\sqrt{2+4\Im(z)^2}\big\}.
\end{array} 
\end{equation}

We have implemented techniques for calculating the branch cuts inherited by functions acting on non-trivial arguments, and extended this to calculate the cuts of expressions and relations of such functions.  The techniques have been gathered together in a {\sc Maple} package, \texttt{BranchCuts} included as part of {\sc Maple 17} and accessed via the \texttt{FunctionAdvisor} tool. Readers with an earlier version can download the code as detailed in Appendix \ref{APP:Maple}.  Both the sets in (\ref{eq:arcsin}) and the visualisation on the right of Figure \ref{fig:arcsin} were produced by the package.  In fact, all the 2d figures in the paper are produced by the package from the output of the branch cut algorithms, while all the 3d figures are numerical plots of either the real or imaginary parts of the expressions in question.  

{\sc Maple}'s \texttt{FunctionAdvisor} is a handbook for special functions, designed to be both human and machine readable, and interactive, processing the output to fit the query, \cite{ChebTerrab2002}.  It covers topics such as symmetries and series expansions with information for almost all of {\sc Maple}'s built in functions.  In {\sc Maple 16} the functionality for branch cut computation was limited.  There existed a table with the defining cuts for most functions in terms of a variable $z$ and if a function was given with a different argument it would return the definitions with $z$ replaced by that argument.  Presenting branch cuts this way could be unintuitive and in some cases incorrect (for example, when the argument induced its own branch cuts these were not returned).  In {\sc Maple 17} queries to \texttt{FunctionAdvisor} on branch cuts use the \texttt{BranchCuts} package discussed in this paper, and additionally, a variety of options are now available for visualising the cuts. 

The primary motivation for the implementation is a wider project at Bath on \textit{simplification}.  The aim is to develop the technology for computer algebra systems  to safely apply identities for multi-valued functions on their single valued counterparts.  The key idea is to decompose the complex domain using cylindrical algebraic decomposition (CAD) according to the branch cuts of the functions involved, so that the truth value of the proposed identity will be constant in each region of the decomposition and hence may be tested by a sample point.  This decomposition approach was introduced in \cite{DF94} with the method using CAD developed in a series of papers;
\cite{BD02, BD03, BBDP04, BPB05, BBDP07, PBD10}. 
Many of the results are summarised in \cite{Phisanbut11} with the current state discussed recently in \cite{DBEW12}

In this paper we discuss the implementation of the techniques in {\sc Maple}, and the issues raised.  We start in Section \ref{SEC:Calc} by giving pseudo-algorithms describing the implementation.  
These can produce sets of cuts which are a superset of the actual branch cuts, that is, some of the cuts produced may not actually correspond to discontinuities of the functions.  This led us to a classification of the different types of output, presented in Section \ref{SEC:Types}.   
While there has been work on calculating branch cuts before, most notably in \cite{DF94}, our work goes much further with the careful description of the algorithms, their output and how it may be classified. 
Finally, in Section \ref{SEC:SA} we consider the use of this work in simplification technology and the effect of the condition that the input to CAD be a semi-algebraic set (list of polynomial equations or inequalities in real variables). Finally, some details on using the actual {\sc Maple} package are provided in Appendix \ref{APP:Maple}.
Although our implementation is in {\sc Maple}, we note that the ideas presented are relevant for any system to compute branch cuts.

\section{Calculating Branch Cuts} \label{SEC:Calc}


\subsection{Moving to real variables}

We first consider representing branch cuts as portions of algebraic curves in two real variables; the real and imaginary parts of a complex variable, $z$.  

\begin{figure}[ht] 
\begin{center}
\includegraphics[width=4.5cm]{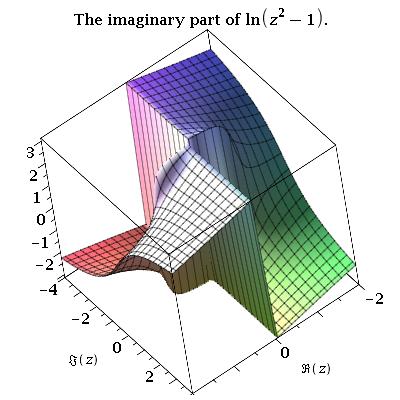}
\hspace*{0.3cm}
\includegraphics[width=4.5cm]{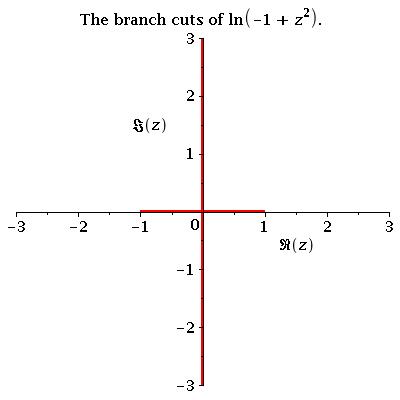}
\end{center}
\caption{Plots relating to $f(z)=\log(z^2-1)$ from Examples \ref{ex:RV} and \ref{ex:CV}.} 
\label{fig:log}
\end{figure}

\begin{example}
\label{ex:RV}
Consider the function $f(z) = \log(z^2-1)$.  The function $\log$ has branch cuts when its argument lies on the negative real axis hence $f(z)$ has branch cuts when $\Im(z^2-1)=0$ and $\Re(z^2-1)<0$.  If we let $x=\Re(z), y=\Im(z)$ then this reduces to 
$2xy = 0, x^2-y^2-1 < 0$,   
with solutions $\{y=0, x \in (-1,1)\}$ and $\{x=0, y \mbox{ free} \}$.  Hence the branch cuts are as shown in Figure \ref{fig:log}.  
\end{example}

This technique is summarised by Algorithm \ref{alg:BC-F-RV1}.  
In the implementation steps 1 and 2 are performed by calls to \texttt{FunctionAdvisor}, accessing the table of defining cuts.  In step 2 we assume that the defining cuts are portions of either the real or imaginary axis encoded as the choice of which is zero and a range over which the other varies.  While not strictly required in theory the assumption is used throughout the implementation.  Then in step 6 the semi-algebraic set will consist of one equality and one or two inequalities (depending on whether the range runs to infinity).  Each solution in step 7 will consist of an equality defining one of $\{x,y\}$ in terms of the other, and a range for the other variable.   Step 7 could be implemented with a variety of techniques.  We use {\sc Maple}'s standard solving tools and find it most efficient to first solve the equality and then consider each possible solution with the inequalities.  In using these tools we are assuming that {\sc Maple} can identify all the solutions, which is not the case for polynomials of high degree.  However, we find them sufficient for all practical examples encountered.

\begin{algorithm}[ht] \caption{BC--F--RV1} \label{alg:BC-F-RV1}
\DontPrintSemicolon
\SetKwInOut{Input}{Input}\SetKwInOut{Output}{Output}
\Input{$f(p(z))$ where $p$ is a polynomial and $f$ has known defining cuts.}
\Output{The branch cuts of the mathematical function defined $f(p(z))$.}
\BlankLine
\eIf{$f$ \rm{introduces branch cuts}}{
Obtain the defining branch cut(s) for $f$. \;
Set $\Re(z)=x, \Im(z)=y$ to obtain $p(z)=p(x,y)$. \;
Set $\mathcal{R}$ and $\mathcal{I}$ to be respectively the real and imaginary parts of $p(x,y)$. \;
\For{ \rm{each defining cut} $C_i$}{
Define a semi-algebraic set in $(x,y)$ by substituting $\mathcal{R}$ and $\mathcal{I}$ into $C_i$.  \;
Set $B_i$ to be the set of solutions to the semi-algebraic set. 
}
\Return The union of the $B_i$. \;
}
{\Return the empty set. \;}
\end{algorithm}

\subsection{Combinations of functions}

We extend Algorithm \ref{alg:BC-F-RV1} to study {\em combinations} of functions (sums, products and relations) by applying the algorithm to each component and then taking the union of the sets of branch cuts in the outputs, as specified in Algorithm \ref{alg:BC-C}.  In step 3 a suitable algorithm is one beginning BC--F that accepts $F_i$ as input.

Note that the output specification of Algorithm \ref{alg:BC-C} is looser than that of Algorithm \ref{alg:BC-F-RV1}.  One reason for this is that a combination of functions with branch cuts may have their individual branch cuts intersecting, and if the discontinuities introduced are equivalent then these would cancel out as in Example \ref{ex:comb}.  In Section \ref{SEC:Types} we classify the output of these algorithms, including output relating to these cancellations, (Definition \ref{def:spurioustypes}).
 
\begin{algorithm}[ht] \caption{BC--C} \label{alg:BC-C}
\DontPrintSemicolon
\SetKwInOut{Input}{Input}\SetKwInOut{Output}{Output}
\Input{Any combination of functions whose branch cuts can individually be studied by an available algorithm.}
\Output{A set of cuts, a subset of which are the branch cuts of the mathematical function defined by the expression.}
\BlankLine
Set $F_1, \dots F_n$ to be the functions involved in the expression. \;
\For{$i = 1 \dots n$}
{Set $B_i$ to the output from applying a suitable branch cuts algorithm to $F_i$.}
\Return $\cup_i B_i$ 
\end{algorithm}

\begin{figure}[ht]
\begin{center}
\includegraphics[width=4.5cm]{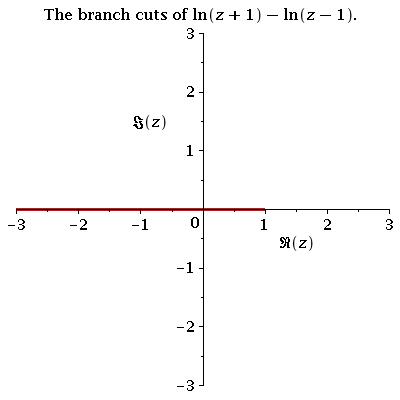}
\hspace*{0.3cm}
\includegraphics[width=4.5cm]{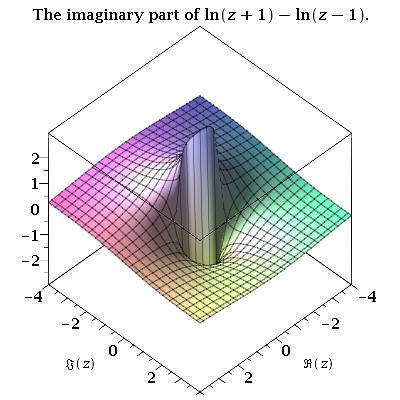}
\end{center}
\caption{Plots relating to $f(z)=\log(z+1) - \log(z-1)$ from Example \ref{ex:comb}.} 
\label{fig:comb}
\end{figure}

\begin{example} \label{ex:comb}
Let $f(z)=\log(z+1) - \log(z-1)$ and use Algorithm \ref{alg:BC-C} to identify the branch cuts.  First we use Algorithm \ref{alg:BC-F-RV1} to identify the branch cut of the first term as the real axis below $-1$ and the branch cut of the second to be the real axis below $1$.  Hence Algorithm \ref{alg:BC-C} returns the union; the real axis below $1$ as visualised on the left of Figure \ref{fig:comb}.  However, the function actually only has discontinuities on the real axis in the range $(-1,1)$ as demonstrated by the plot on the right of Figure \ref{fig:comb}.  Crossing the negative real axis below $-1$ does induce a discontinuity in the imaginary part of both terms.  However, those discontinuities are equal and so cancel each other out in the expression for $f(z)$.
\end{example}

\subsection{Allowing nested roots}

We can extend Algorithm \ref{alg:BC-F-RV1} to let $p$ be a rational function by modifying step 7 to multiply up after substituting $\mathcal{R}$ and $\mathcal{I}$ into $C_i$.  The question of zero denominators will only arise if the input $p$ itself has a zero denominator and so we might assume this issue would have been dealt with previously.

We can relax the input specification further by allowing nested roots, more specifically, by letting the argument belong to the class of radical expressions in $z$ (expressions built up from $+,-,/,*$ and $\sqrt[n]{}$ where $n$ is a natural number greater than 1).  This is because such an argument can be modified to give a rational function from which information on the real and imaginary parts of the original argument can be inferred, a process known as \textit{de-nesting} the roots.  Hence we can still obtain a semi-algebraic set representing the branch cuts as before.  

By de-nesting the roots we may end up with extra solutions which do not define branch cuts of the input function.  For example, consider a function with argument $q(z)$ which when squared gives $q^2=p(z)$, a rational function in $z$.  However, this now represents the solution set $q(z) = \pm p(z)$, i.e. solutions for both branches of the square root, instead of just the desired principal branch.  Ideally these erroneous solutions should be identified and removed.

Another issue in relaxing the input specification is that we must now consider the possibility of extra branch cuts arising from the argument itself.  Taking these issues into account, we describe Algorithm \ref{alg:BC-F-RV2}.  This is a modification of Algorithm \ref{alg:BC-F-RV1} with a relaxed input specification, leading to looser output specification.  

\begin{algorithm}[ht] \caption{BC--F--RV2} \label{alg:BC-F-RV2}
\DontPrintSemicolon
\SetKwInOut{Input}{Input}\SetKwInOut{Output}{Output}
\Input{$f(q(z))$ where $q$ is a radical expression and $f$ has known defining cuts.}
\Output{A set of cuts, a subset of which are the branch cuts of the mathematical function defined by $f(q(z))$.}
\BlankLine
\eIf{$f$ \rm{introduces branch cuts}}{
Obtain the defining branch cuts for $f$. \;
Set $z=x+iy$ to obtain $q(z)=q(x,y)$. \;
De-nest the roots in $q(x,y)$ to obtain $p(x,y)$.
Set $\mathcal{R}_p$ and $\mathcal{I}_p$ to be respectively the real and imaginary parts of $p(x,y)$. \;
Define a semi-algebraic set in $(x,y)$ from $\mathcal{R}_p$ and $\mathcal{I}_p$ using information from the defining cuts.  \;
Set $B$ to be the solutions of the semi-algebraic set.  \;
If possible, remove erroneous solutions arising from the de-nesting. \;
}
{Set $B$ to be the empty set. \;}
Set $A = $BC--C($q(z)$) (application of Algorithm \ref{alg:BC-C}). \;
\Return $A \cup B$.
\end{algorithm}

Various methods for de-nesting roots and removing the erroneous solutions have been studied in
\cite{BBDP04, BPB05, BBDP07, Phisanbut11}.  
The \texttt{BranchCuts} package currently has only a very limited implementation of the squaring method outlined above, but further work is planned.  Note that even this simple implementation can induce the erroneous solutions discussed as outlined by Example \ref{ex:problem}.

\begin{figure}[ht] 
\begin{center}
\includegraphics[width=4.5cm]{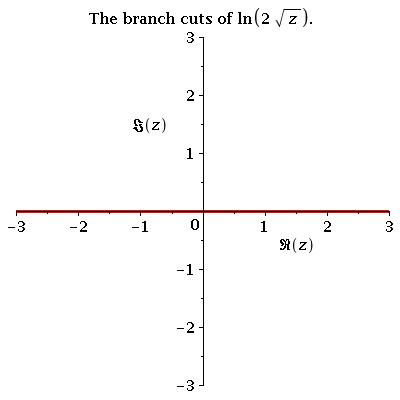}
\hspace*{0.3cm}
\includegraphics[width=4.5cm]{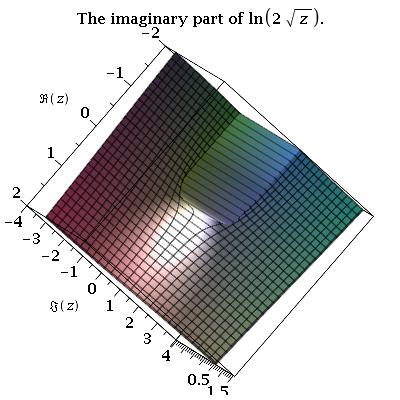}
\end{center}
\caption{Plots relating to $f(z)=\log(2\sqrt{z})$ from Example \ref{ex:problem}.} 
\label{fig:problem}
\end{figure}

\begin{example} \label{ex:problem}
Let $f=\log(2\sqrt{z})$ and use Algorithm \ref{alg:BC-F-RV2} to identify the branch cuts.  First we set $q=2\sqrt{z}=2\sqrt{x+iy}$.  Then we de-nest by squaring to give $p=q^2=4(x+iy)$.  In this simple example, 
\begin{equation} 
\label{eq:simpleEx1}
\mathcal{R}_p=4x \qquad \mbox{and} \qquad \mathcal{I}_p=4y.
\end{equation}
We suppose that $q = \mathcal{R}_q+\mathcal{I}_q i$ and hence
\begin{equation}
\label{eq:simpleEx2}
p = \mathcal{R}_q^2-\mathcal{I}_q^2 + 2i(\mathcal{R}_q\mathcal{I}_q)
\end{equation}
Since $\log$ has defining cuts along the negative real axis we know $\mathcal{R}_q<0$ and $\mathcal{I}_q=0$.  Upon comparing (\ref{eq:simpleEx1}) and  (\ref{eq:simpleEx2}) we see the second condition implies $y=0$ and $x=\frac{1}{4}\mathcal{R}_q^2 >0$.  In this example the first condition offers no further information (if the defining cut had not run to $\infty$ it could have bounded $x$).  Hence we define the semi-algebraic set $\{y=0, x>0\}$.  
We also compute the set $\{y=0, x<0\}$, which is the branch cut of $q(z)$ itself, and return the union of the sets which together specify the entire real axis as presented visually on the left of Figure \ref{fig:problem}.  

Unfortunately, the function only actually has discontinuities over the negative real axis, as demonstrated by the plot of the right of Figure \ref{fig:problem}.  The first solution set was erroneous.  This is clear since if $x>0$ and $y=0$ then $\sqrt{z}>0$ and so can never lie on the negative real axis.  The solution related to the case $q=-\sqrt{p}$ which was not relevant to the problem.  
(The reason for the factor of $2$ in the example is because {\sc Maple} automatically simplifies $\log(\sqrt{z})$ to $\frac{1}{2}\log(z)$ which can be analysed by Algorithm \ref{alg:BC-F-RV2} to give exactly the branch cuts of the function.)
\end{example}

\subsection{Using a complex parametric representation}

We now consider a second approach to representing branch cuts, first suggested by \cite{DF94}.  Rather than moving to real variables this approach defines cuts using a complex function of a real parameter and a range for that parameter.

\begin{example}
\label{ex:CV}
Let $f(q(z)) = \log(z^2-1)$, the function from Example \ref{ex:RV}.   We consider when $q$ takes values on the branch cuts of $f$ by setting $q=a$ where $a$ ranges over the cuts.  In this case $z^2-1=a$ can be easily rearranged to give $z(a)=\pm \sqrt{a+1}$.  Hence we can represent the branch cuts by the two roots, each presented with the range $a \in (-\infty, 0)$.  By considering the behaviour of the functions for various portions of the parameter range we see that these define the same cuts as presented in Example \ref{ex:RV} and visualised on the right of Figure \ref{fig:log}.
\end{example}

This technique is summarised by Algorithm \ref{alg:BC-F-CV}.  In this case the 
assumption that the defining cuts are portions of either the real or imaginary axis is really required.   If $q(z)$ is a radical expression containing nested roots then step 5 will require de-nesting and so the output may be a superset of the actual branch cuts.  (For example, the set produced for $\log(2\sqrt{z})$ is equivalent to that produced in Example \ref{ex:problem}.)  Note that Algorithm \ref{alg:BC-F-CV} could have been provided with the input and output specifications of Algorithm \ref{alg:BC-F-RV2} (i.e if $q(z)$ were a radical expression then given sufficient computing resources all the branch cuts could be identified as part of a larger set).  Instead we have provided the specifications used for the implementation.  This does not restrict the possibilities for $q(z)$, instead building in a warning system to ensure the correctness of the output.  In particular this allows $q$ to contain any elementary function, returning not the complete (possibly infinite) set of branch cuts but at least those in the principal domain.

\begin{algorithm}[ht] \caption{BC--F--CV} \label{alg:BC-F-CV}
\DontPrintSemicolon
\SetKwInOut{Input}{Input}\SetKwInOut{Output}{Output}
\Input{$f(q(z))$ where $f$ has known defining cuts.}
\Output{A set of cuts which either contain the branch cuts of the mathematical function defined by $f(q(z))$ as a subset, or are accompanied with a warning that this is not the case.}
\BlankLine
\eIf{$f$ \rm{introduces branch cuts}}{
Obtain the defining branch cuts for $f$, each a range on an axis.  \;
\For{ \rm{each cut} $C_i$}{
If $C_i$ is on the real axis then set $q(z) = a$, otherwise set $q(z)=ia$. \;
Find the solutions $z(a)$ to this equation. If the complete set of solutions cannot be guaranteed then provide a \textbf{warning}\;
Set $B$ to be the set of solutions, each given with the range for $a$ from $C_i$. \;
If possible, remove erroneous solutions arising from any de-nesting. \;
}
}
{Set $B$ to be the empty set. \;}
Set $A = $BC--C($q(z)$) (application of Algorithm \ref{alg:BC-C}).\;
\Return $A \cup B$.
\end{algorithm}

This approach was simple to implement in {\sc Maple} using the \texttt{solve} command as a black box for step 5.  (As discussed before Algorithm \ref{alg:BC-F-RV1}, this is making assumptions on the solve tools which would not always be valid, but they are found to be sufficient for all practical examples encountered.)   
The main advantage of this approach over moving to real variables is that it tends to be much quicker, especially when there are nested roots.  The major disadvantage is that the output is usually far more complicated (requires much more space to display), often contains components that map to the same cuts, and is far less intuitive (the curves encoded are not visible algebraically).  Example \ref{ex:Kahan} demonstrates some of these features.  Despite the often unwieldy output, {\sc Maple}'s plotting features allows for the position and nature of the cuts to be quickly made apparent.

For these reasons it is expected that the earlier algorithms are more useful for implementation in other code and use in mathematical study while Algorithm \ref{alg:BC-F-CV} is very useful for getting a quick visual understanding of the branch cuts in an expression and may have much utility in practical applications for this purpose. 

\begin{figure}[ht] 
\begin{center}
\includegraphics[width=4.5cm]{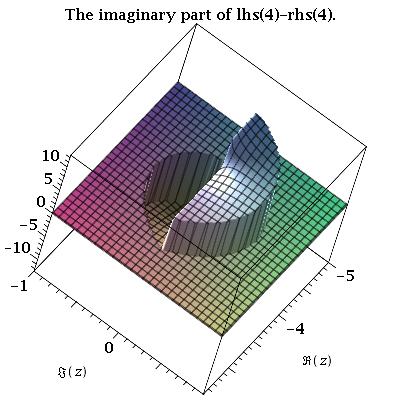}
\hspace*{0.3cm}
\includegraphics[width=4.5cm]{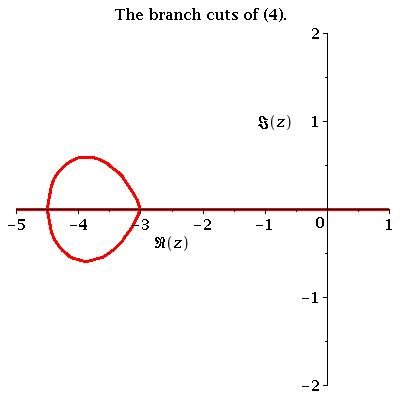}
\end{center}
\caption{Plots relating to equation (\ref{eq:Kahan}) from Example \ref{ex:Kahan}.} 
\label{fig:Kahan}
\end{figure}

\begin{example}
\label{ex:Kahan}
A classic example within the theory of branch cut calculation and simplification is that of Kahan's teardrop, from \cite{Kahan87}.  Kahan considers the relation
\begin{equation}
\label{eq:Kahan}
2\rm{arccosh}\left(\frac{3+2z}3\right)-\rm{arccosh}\left(\frac{5z+12}{3(z+4)}\right)= 2\rm{arccosh}\left(2(z+3)\sqrt{\frac{z+3}{27(z+4)}}\right)
\end{equation}
noting that it is true for all values of $z$ in the complex plane except for a small teardrop shaped region over the negative real axis, as demonstrated by the plot on the left of Figure \ref{fig:Kahan}.  Both of the approaches to calculating branch cuts outlined above will return a set represented visually by the image on the right of Figure \ref{fig:Kahan}.  However, the algebraic representations are quite different.  When working in real variables the upper half of the teardrop is represented by the set
\[
\left\{ \Im(z) = \frac{ \sqrt{-(2\Re(z)+5)(2\Re(z)+9)}(\Re(z)+3) }{2\Re(z)+5}, \, -\frac{5}{2} < \Re(z), \Re(z) < -\frac{9}{4} \right\}
\]
while using the complex parametric approach the same portion of the teardrop is given by  
\begin{eqnarray*}
z &=& \frac{-3}{4\left( 2\,a-{a}^{3}+2\,\sqrt {-{a}^{2} \left( -1+{a}^{2} \right) }  \right) ^{2/3}}
\left[ \left( 2\,a-{a}^{3}+2\,\sqrt {-{a}^{2} \left( -1+{a}^{2} \right) } \right) ^{2/3} \right. \\
& & \quad 
-3\,i\sqrt {3}{a}^{2}  -3\,ia\sqrt {3}\sqrt {-{a}^{2} \left( -1+{a}^{2} \right) }
-3\,\sqrt [3]{2\,a-{a}^{3}+2\,\sqrt {-{a}^{2} \left( -1+{a}^{2} \right) }}a \\
& & \quad -3\,\sqrt [3]{2\,a-{a}^{3}+2\,
\sqrt {-{a}^{2} \left( -1+{a}^{2} \right) }}\sqrt {-{a}^{2} \left( -1+{a}^{2} \right) }-3\,{a}^{2}
\\
\end{eqnarray*}
\begin{eqnarray*}
& & \quad -3\,a\sqrt {-{a}^{2} \left( -1+{a}^{2} \right) }
+3\,i\sqrt [3]{2\,a-{a}^{3}+2\,\sqrt {-{a}^{2} \left( -1+{a}^{2} \right) }}\sqrt {3}a\\
& & \left. \quad +3\,i\sqrt [3]{2\,a-{a}^{3}+2\,\sqrt {-{a}^{2} \left( -1+{a}^{2} \right) }}\sqrt {3}\sqrt {-{a}^{2} \left( -1+{a}^{2} \right)  } \right]
\end{eqnarray*}
with $a$ running over the range $(-1,1)$.  
\end{example}

We note that the identity (\ref{eq:Kahan}) in Example \ref{ex:Kahan} is actually introduced by a fluid mechanics problem and so this example demonstrates how issues relating to branch cuts may be encountered by users of multi-valued functions in other fields.  Hence the importance of understanding them fully and the benefit of an accurate and intuitive representation.

\section{Classification of Branch Cuts} 
\label{SEC:Types}

The work of Section \ref{SEC:Calc} raises several issues and necessitates a classification of the different cuts that can be produced by these methods.  It is common to use the generic term \textbf{branch cuts} to refer to any curve portions that are defining cuts of functions or output from the algorithms.  We classify these, starting with the definition most usually meant by users.

\begin{definition}
Let $F$ be an analytic multi-valued function and $f$ its single-valued counterpart.  The \textbf{branch cuts of the mathematical function} (called \textbf{true cuts} for brevity) are the curves or curve segments over which $f$ is discontinuous, corresponding to $F$ moving to another branch of its Riemann surface.  
\end{definition}

Hence all the defining branch cuts are true cuts, as are any cuts produced by Algorithm \ref{alg:BC-F-RV1}.  However, as demonstrated by Examples \ref{ex:comb} and \ref{ex:problem} the other algorithms may give output that does not adhere to this definition.

\begin{definition}
Define any branch cuts calculated by the algorithms over which the function is actually continuous as \textbf{spurious cuts}.
\end{definition}

(In \cite{DF94} the authors used the term {\em removable} instead of spurious, in analogy with removable singularities.)  All branch cuts may be classified as either true or spurious.  We further classify spurious cuts according to their origin.

\begin{definition}
\label{def:spurioustypes}
Define those spurious cuts introduced through a de-nesting procedure as \textbf{de-nesting cuts}, while those introduced by the intersection of true cuts from different parts of an expression as \textbf{formulation cuts}.
\end{definition}

Hence all spurious branch cuts produced by the algorithms in this paper are either de-nesting cuts or formulation cuts.  Some spurious cuts may be both (or more accurately there may be two cuts, one of each type, which coincide).

Note that the output of Algorithms \ref{alg:BC-F-RV2} and \ref{alg:BC-F-CV} are collections of true cuts and de-nesting cuts, while the output of Algorithm \ref{alg:BC-C} is a collection of true cuts, de-nesting cuts and formulation cuts. 

\begin{figure}[ht] 
\begin{center}
\includegraphics[width=3.7cm]{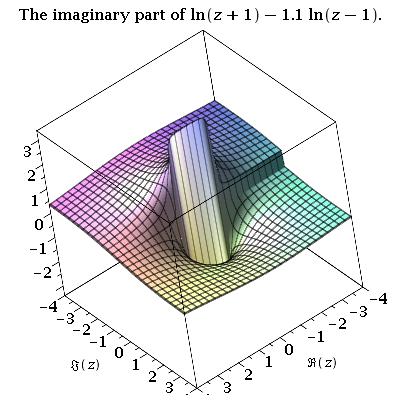}
\hspace*{0.3cm}
\includegraphics[width=3.7cm]{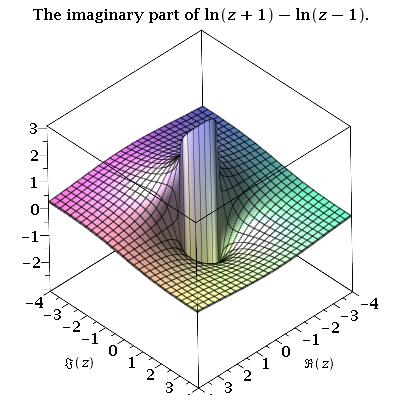}
\hspace*{0.3cm}
\includegraphics[width=3.7cm]{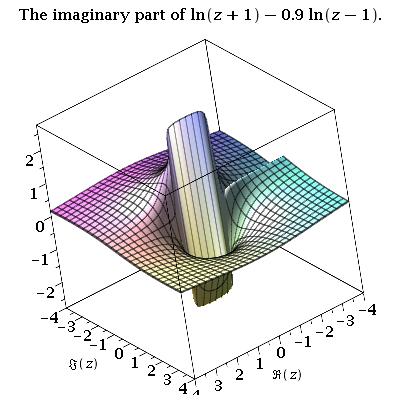}
\end{center}
\caption{Plots relating to $f_{\epsilon}(z)=\log(z+1) - \epsilon\log(z-1)$.} 
\label{fig:Inherent}
\end{figure}

It would be desirable to have algorithms to remove all spurious cuts, or perhaps better, algorithms that do not introduce them in the first place.  There has already been work on the removal of certain spurious cuts in \cite{DF94} and \cite[etc.]{Phisanbut11} and this will be the topic of more study.  We feel that formulation cuts will be the more difficult to avoid since they are inherent to the formulation of the mathematical function chosen in the expression given to an algorithm.  Consider 
\[
f_{\epsilon}(z)=\log(z+1) - \epsilon\log(z-1).
\]
When $\epsilon=1$ we are in the case of Example \ref{ex:comb} and applying Algorithm \ref{alg:BC-C} will result in a branch cut over the real axis below $1$, with the portion between $-1$ and $1$ being a true cut and the rest a spurious cut.  However, if we let $\epsilon$ differ at all from $1$ then the spurious cuts will instantly become true.  The size of the discontinuities will depend on the magnitude of $\epsilon$ but their presence does not.  Figure \ref{fig:Inherent} shows the presence of the true cuts occurring when $\epsilon$ varies by just one tenth from $1$.  

\section{Semi-algebraic Output for Simplification Technology} \label{SEC:SA}

As discussed in the introduction, the primary motivation for this work was application in simplification technology, based on decomposing the domain according to the branch cuts of proposed simplifications using CAD.  However, most CAD algorithms require the input to be a semi-algebraic set (list of polynomial equations and inequalities), with the polynomials usually defined over the field of rational coefficients.  None of the algorithms described so far give such output, however Algorithms \ref{alg:BC-F-RV1} and \ref{alg:BC-F-RV2} could be easily modified to do so, by terminating early and returning the output of steps 5 and 6 respectively.  We denote such an algorithm by BC$-$F$-$SA and note that it could be used on combinations via Algorithm \ref{alg:BC-C}.  
For Example \ref{ex:RV} BC$-$F$-$SA would return $\{2xy = 0, x^2-y^2-1 < 0\}$.  However, for more complicated examples, the output may contain far more information than required to describe the cuts.

\begin{figure}[ht] 
\begin{center}
\includegraphics[width=3.7cm]{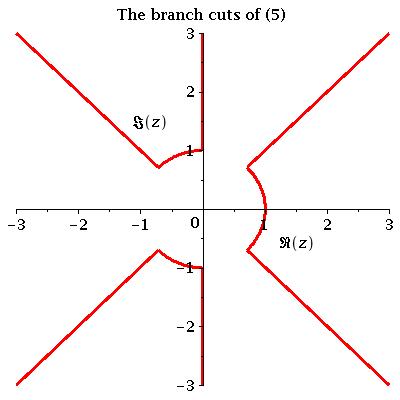}
\hspace*{0.3cm}
\includegraphics[width=3.7cm]{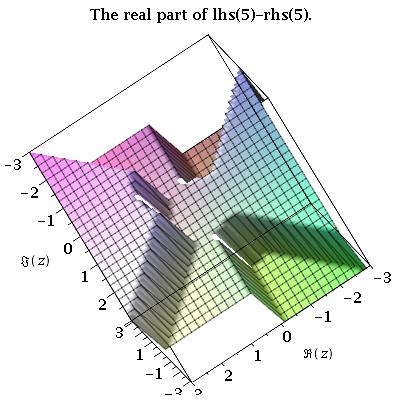}
\hspace*{0.3cm}
\includegraphics[width=3.7cm]{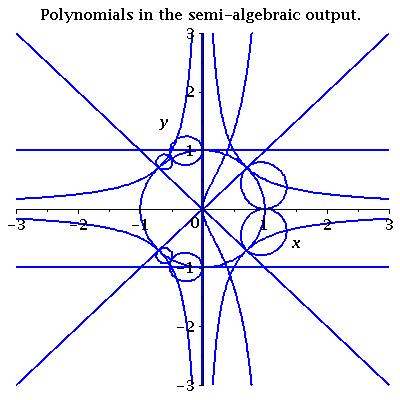}
\end{center}
\caption{Plots relating to equation (\ref{eq:arctan}) from Example \ref{ex:SA}.} 
\label{fig:SA}
\end{figure}

\begin{example}
\label{ex:SA}
Consider the formula 
\begin{equation}
\label{eq:arctan}
\arctan(z) + \arctan(z^2) = \arctan \left( \frac{z(1+z)}{(1-z^3)} \right).
\end{equation}
The plot on the left of Figure \ref{fig:SA} is a visualisation for the output from either of the approaches in \texttt{BranchCuts}, while the true cuts are apparent from the centre plot.
Define, 
\begin{align*} 
f(x,y) &= \left( 1-x \right) {y}^{4} - \left( 2\,{x}^{3}+1 \right) {y}^{2} - {x}^{5} - {x}^{4} + {x}^{2} + x \\
g(x,y) &= {y}^{6} - {y}^{5} + 3\,{x}^{2}{y}^{4} - 2\left( {x}^{2} + x \right){y}^{3} 
\\ &\qquad 
+ 3\left( 2\,x+{x}^{4} \right){y}^{2} - \left( {x}^{4} + 2\,{x}^{3} +2\,x + 1 \right)y + {x}^{6} - 2\,{x}^{3}.
\end{align*}
If we were to instead take the semi-algebraic output, then we would have the following list of semi-algebraic sets,
\begin{align*}
&\{x=0,-y\leq -1\}, \{ {x}^{2}-{y}^{2}=0,-2\,xy\leq -1\}, \{f(x,y)=0,  g(x,y) \leq -1\},  \\
&\{x=0, y\leq -1\}, \{ {x}^{2}-{y}^{2}=0, 2\,xy\leq -1\}, \{f(x,y)=0, g(x,-y) \leq -1\}.
\end{align*}
A full sign-invariant CAD for this problem would ignore the relation signs and just consider the polynomials present.  A plot of the polynomials is given on the right of Figure \ref{fig:SA} and clearly contains far more information than required to understand the branch cuts.  Note that the correctness of the original formula is governed only by the branch cuts and hence the plot on the left: equation (\ref{eq:arctan}) is true in the connected region containing the origin and false in the other three full dimensional regions.  A CAD allows us to find the regions of truth and falsity axiomatically by testing each cell of the CAD using a sample point.
\end{example}

There are various smarter approaches than calculating a full sign-invariant CAD, such as partial CADs and equational constraints.  Work on a CAD based method that can take into account more of the structure of problems of branch cut analysis has recently been reported in \cite{BDEMW13}, and studied further in \cite{BDEW13}.

\section{Summary and Future Work}

We have considered the problem of calculating the branch cuts of expressions, presenting two approaches and describing their implementation as part of {\sc Maple} 17.  We have classified the output of our algorithms and described how they could be adapted to provide semi-algebraic output for simplification technology.  We are currently working on developing such technology based on the new concept of a truth table invariant CAD \cite{BDEMW13, BDEW13}; a decomposition which can be more closely fitted to the semi-algebraic description of branch cuts.  

Future work with branch cuts will include the generalisation to many complex variables and the utilisation of better knowledge of branch cuts elsewhere, such as for choosing intelligent plot domains.  Most importantly will be the further characterisation of spurious cuts and methods to remove them from the output.

\subsection*{Acknowledgements}

This work was supported by the EPSRC grant: EP/J003247/1.  The code in {\sc Maple} 17 is a collaboration between the University of Bath and Maplesoft.  We would particularly like to thank Edgardo Cheb-Terrab from Maplesoft for his interest in the work and contribution to the code.

\appendix

\section{The \texttt{BranchCuts} Package in {\sc Maple 17}} \label{APP:Maple}

The \texttt{BranchCuts} package is part of the \texttt{MathematicalFunctions} package in {\sc Maple 17}, but is usually accessed directly by queries to the \texttt{FunctionAdvisor}.  To access the commands individually in Maple 17 use \\
\texttt{> kernelopts(opaquemodules=false):} \\ 
\texttt{> with(MathematicalFunctions:-BranchCuts):} \\
Readers with an earlier version of {\sc Maple} can download a file with the code from \texttt{http://opus.bath.ac.uk/32511/} along with an introductory  worksheet demonstrating its installation and use. 

Two key commands are available; \texttt{BCCalc} which produces branch cuts using the algorithms of this paper and \texttt{BCPlot} which can make 2d visualisations of the output.  There are two mandatory arguments for \texttt{BCCalc}; the expression to be considered and the variable.  The key optional argument is the choice of method.  Providing \texttt{method=RealVariables} will cause \texttt{BCCalc} to use Algorithms \ref{alg:BC-F-RV1} and \ref{alg:BC-F-RV2} while providing \texttt{method=ComplexVariable} will use Algorithm \ref{alg:BC-F-CV}.  The default, chosen for efficiency, uses Algorithm \ref{alg:BC-F-RV1} where possible and Algorithm \ref{alg:BC-F-CV} elsewhere.  Combinations of functions are dealt with using Algorithm \ref{alg:BC-C}.

\begin{figure}[hb] 
\begin{center}
\includegraphics[width=3.7cm]{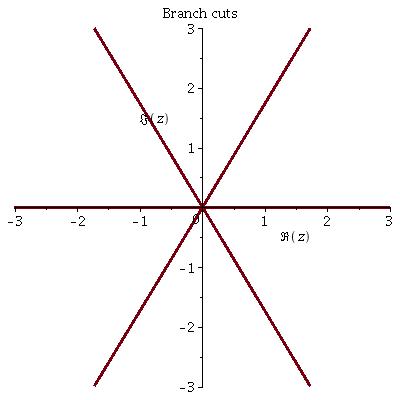}
\hspace*{0.3cm}
\includegraphics[width=3.7cm]{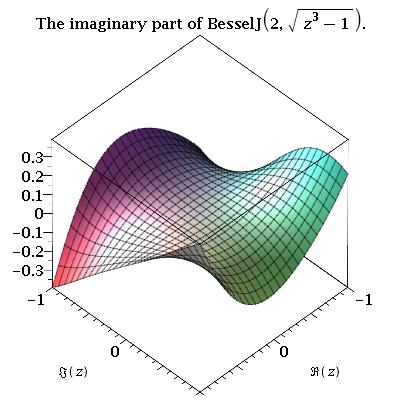}
\hspace*{0.3cm}
\includegraphics[width=3.7cm]{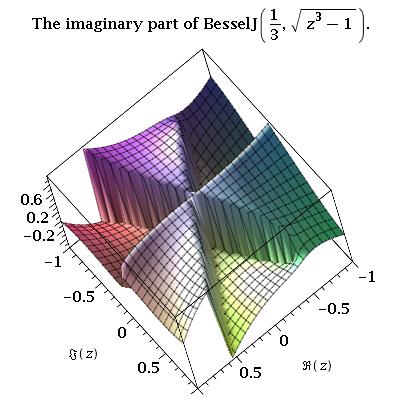}
\end{center}
\caption{Plots relating to BesselJ$(a,\sqrt{z^3-1})$.} 
\label{fig:Bessel}
\end{figure}

The specification of the algorithms are checked but not strictly enforced.  Instead warnings are provided if the method is not applicable or the output cannot be guaranteed to contain all true cuts.  The package can work with any function whose defining cuts (or lack of cuts) is recorded in the \texttt{FunctionAdvisor} table.  It covers all elementary functions and many others such as Bessel functions, Jacobi $\theta$-functions and Chebyshev polynomials.  These examples are actually multivariate in a computer algebra sense (univariate functions with parameters in a mathematical sense).  Their branch cuts can be considered since they only occur with respect to one variable.  If the presence of the branch cuts depends on the value of the parameters then the condition is checked.  If it cannot be determined true or false (say if the relevant parameter has not been set), then the branch cut is included but a relevant warning is given.  
For example,
\begin{verbatim}
> BCCalc( BesselJ(a,sqrt(z^3-1)), z, 
          parameters={a}, method=RealVariables );
\end{verbatim}
produces the message, \verb+Warning, branch cuts have been calculated which+ \verb+only occur if a::(Not(integer))+,
and outputs the six branch cuts
\begin{eqnarray*}
&\{\Im(z) = 0, \Re(z) < 1\}, \, \{\Im(z) = 0, 1 < \Re(z)\}, \\
&\{\Re(z) = -\frac{1}{3}\sqrt{3}\Im(z), \frac{1}{2}\sqrt{3} < \Im(z)\}, \,
\{\Re(z) = -\frac{1}{3}\sqrt{3}\Im(z), \Im(z) < \frac{1}{2}\sqrt{3}\}, \\
&\{\Re(z) = \frac{1}{3}\sqrt{3}\Im(z), -\frac{1}{2}\sqrt{3} < \Im(z)\}, \,
\{\Re(z) = \frac{1}{3}\sqrt{3}\Im(z), \Im(z) < -\frac{1}{2}\sqrt{3}\}.
\end{eqnarray*}
Applying \texttt{BCPlot} to this output produces the image on the left of Figure \ref{fig:Bessel}.  The true cuts for two specific values of $a$ can be observed in the centre and right plots, demonstrating the validity of the warning message.

\end{document}